# Observation of large spontaneous emission rate enhancement of quantum dots in a broken-symmetry slow-light waveguide

Hamidreza Siampour,[1,§,*] Christopher O'Rourke,[1] Alistair J. Brash,[1] Maxim N. Makhonin,[1] René Dost,[1] Dominic J. Hallett,[1] Edmund Clarke,[2] Pallavi K. Patil,[2] Maurice S. Skolnick,[1] A. Mark Fox[1]

[1] Department of Physics and Astronomy, University of Sheffield, Hicks Building, Sheffield S3 7RH, UK

[2] EPSRC National Epitaxy Facility, Department of Electronic and Electrical Engineering, University of Sheffield, Sheffield S1 3JD, UK

[§] Current address: Cavendish Laboratory, University of Cambridge, J. J. Thomson Avenue, Cambridge CB3 0HE, UK

[*] Email: hs807@cam.ac.uk

**Abstract-** Quantum states of light and matter can be manipulated on the nanoscale to provide a technological resource for aiding the implementation of scalable photonic quantum technologies[1-3]. Experimental progress relies on the quality and efficiency of the coupling between photons and internal states of quantum emitters[4-6]. Here we demonstrate a nanophotonic waveguide platform with embedded quantum dots (QDs) that enables both Purcell-enhanced emission and strong chiral coupling. The design uses slow-light effects in a glide-plane photonic crystal waveguide with QD tuning to match the emission frequency to the slow-light region. Simulations were used to map the chirality and Purcell enhancement depending on the position of a dipole emitter relative to the air holes. The highest Purcell factors and chirality occur in separate regions, but there is still a significant area where high values of both can be obtained. Based on this, we first demonstrate a record large radiative decay rate of 17±2 ns$^{-1}$ (60±6 ps lifetime) corresponding to a 20±2 fold Purcell enhancement. This was achieved by electric-field tuning of the QD to the slow-light region and quasi-resonant phonon-sideband excitation. We then demonstrate a 5±1 fold Purcell enhancement for a dot with high degree of chiral coupling to waveguide modes, substantially surpassing all previous measurements. Together these demonstrate the excellent prospects for using QDs in scalable implementations of on-chip spin-photonics relying on chiral quantum optics.

**Introduction**

Purcell recognised that the spontaneous emission rate of a quantum system can be modified by its environment[7]. In particular, enhanced photon emission rates have been reported by coupling quantum emitters to high Q-factor micro- and nano-cavities[8-10], dielectric nanoantennas[11] and plasmonic nanostructures[12-16], leading to efficient and bright single-photon sources with the potential for high repetition rates. Moreover, the fast radiative rate helps to mitigate the effects of dephasing processes in solid-state systems, as required for the generation of indistinguishable photons[17].

The benefits of Purcell-enhanced emission can be brought to chip-based platforms by integrating the quantum emitter into photonic-crystal waveguides (PCWs)[18,19]. When optimised, the waveguides enable highly efficient photon extraction into scalable planar architectures. The Purcell enhancement factor $F_P$ in a waveguide is proportional to the group index $n_g = c/v_g$, instead of the cavity Q-factor. Here, the group velocity $v_g$ is the inverse of the first order dispersion $(dk/d\omega)^{-1}$, where $k$ is the wavenumber, and $\omega$ is the angular frequency, and $c$ is the speed of light *in vacuo*. The group velocity is strongly reduced in the slow-light region near the photonic band edge[20], leading to large values of $n_g$, and hence large Purcell factors. Following this approach, emission rates of up to 6.28 $ns^{-1}$ (corresponding to a 9-fold Purcell factor) have been reported for quantum dots (QDs) embedded within a W1 waveguide formed when one row of air holes is omitted from a triangular lattice[17,18]. In this regime, a high coupling efficiency of $\beta$ =98.4% was reported which corresponds to a cooperativity of $\eta$ = 62.7 for the QD-waveguide system[18]. The large emitter-waveguide cooperativity is the key ingredient to scale up the system and coherently couple multiple emitters. Other types of waveguide designs have also been considered, and a Purcell enhancement of up to 12-fold was very recently reported for a QD embedded in a topological slow-light valley PCW formed in a lattice of triangular nanoholes with a bearded interface[21].

A waveguide geometry offers additional benefits by enabling chiral coupling between the spin of the quantum emitter and the helicity of the waveguide mode. These effects exploit the confinement of the light on nano-photonic scales in such a way that a circular dipole positioned at a chiral (C) point couples to left- or right-propagating modes with a near-unity directionality according to its helicity. The circular selection rules of QD exciton transitions then provide a mechanism for directional coupling to internal QD spin states, pushing the system into the regime of chiral quantum optics[22] and opening possibilities for developing spin-photon networks for applications in scalable on-chip quantum information-processing[23,24]. Such chiral coupling effects have been observed for QDs in several different types of nano-photonic waveguides, notably nanobeams[25], glide-plane PCWs[24,26], and topological PCWs[27].

The full advantages of chiral quantum optics are only achieved for QDs that experience strong chiral coupling and high Purcell enhancement at the same time. However, in all of the devices studied so far, the maximal spin dependent directionality occurs for QDs located at chiral points displaced from the centre of the waveguide in the areas of low field intensity, leading to poor emitter-waveguide coupling and therefore low Purcell enhancement. Glide-plane symmetry PCWs have been proposed and engineered to combine efficient chiral coupling with strong Purcell enhancement[24,26] as required for example for effective quantum networks, but experimental realizations of these structures have so far only achieved a limited 2-fold Purcell enhancement factor[28]. The important goal of strong chiral coupling with a high Purcell factor thus remains a significant experimental challenge.

In this paper we report on glide-plane PCW devices in which we have observed record high Purcell factors for both non-chirally-coupled and chirally-coupled quantum dots emitting in the slow-light spectral region of a glide-plane line defect PCW. We achieve this by careful tuning of the QD frequency relative to the photonic band edge using both electric and magnetic fields. In addition, for the non-chirally-coupled dot we used a quasi-resonant phonon-sideband excitation scheme to eliminate slow internal relaxation and hence achieve the true Purcell-enhanced emission rate. In this way we observe a non-chirally-coupled dot with an extremely large radiative decay rate of 17±2 ns$^{-1}$ (60±6 ps lifetime), corresponding to a 20±2 fold Purcell enhancement. We also observe chiral routing of spin-carrying photons from a QD near a C-point with a 5±1 fold Purcell factor, substantially outperforming previous demonstrations with nanobeam[25], nanofiber[29], and PCWs[24,28,30-34]. Together, these results constitute significant progress towards reaching the threshold for scalable chiral networks.

**Sample design and characterization**

Figure 1a shows a schematic of the device layout with PCW structures patterned on a gallium arsenide (GaAs) wafer. Figure 1b shows a scanning electron micrograph (SEM) of the fabricated waveguide structure with self-assembled QDs located in the centre of the GaAs membrane. The central part of the structure (blue-grey part) is the slow-light section that is connected to the standard nanobeam waveguides through slow-to-fast adapting sections (yellow-grey parts), as described in Supplementary, Section A. The waveguides are terminated with grating outcouplers to enable off-chip coupling.

Figure 1c illustrates the diode structure with embedded InGaAs QDs, and electrical contacts made to the p- and n-GaAs layers. The electrical control allows spectral tuning of the QDs in the slow-light region of the PCW structure with the approximate total tuning range being 2.7 nm. A Faraday geometry magnetic field of up to 4.5 T can be applied along the growth ($z$) axis of the QDs (i.e. out of plane). The B-field lifts the energy degeneracy of the $\sigma^+$ and $\sigma^-$ Zeeman components, enabling measurement of chirality. The B-field also provides additional tuning of the QD lines. The sample was held at 4.2 K in a liquid helium bath cryostat, with the superconducting magnet located in the lower section. A schematic of the optical setup and the cryostat is shown in Figure S5 in Supplementary Section B.

The waveguides were designed with up-down glide-plane symmetry as shown in Figure 1d. This glide-symmetric line defect is a compact form of two parallel line defects with up-down symmetries (so called W1 waveguides) as discussed in Supplementary Section A. As a consequence of two parallel propagating modes, two dispersion curves in the bandgap of the PCW appear and form a protected slow-light region, with approximately a doubling of the bandwidth compared to the standard W1 waveguide and the closing of the gap so that no sharp cut-off occurs (see Supplementary Figure S1). Figure 1e illustrates the simulated band structure of the central glide-plane PCW in which the slow-light region is spectrally broadened through the two dispersion curves that are crossing at the band-edge. The details of design and modelling of the photonic crystal structure and the waveguide adaptor are in Supplementary Section A.

The emitter's decay rate coupled to the waveguide mode ($\Gamma_{wg}$) with longitudinal direction of $x$ can be calculated as[35,36],

$$\Gamma_{wg}/\Gamma_0 = [3\pi c\varepsilon_0|\boldsymbol{E}(x,y).\hat{\boldsymbol{n}}_D|^2]/[2k_0^2 \iint S_x(y,z)dA], \quad (1)$$

where $\boldsymbol{E}(x,y)$ is the electric field profile associated with the guided mode, $\hat{\boldsymbol{n}}_D$ is a unit vector parallel to the transition dipole moment for an emitter embedded in the waveguide at position $(x,y)$, $c$ is the speed of light in vacuum and $k_0$ denotes the propagation constant in free space. $\Gamma_0$ represents the spontaneous emission decay rate in vacuum and $S_x$ denotes the x component of the average of the instantaneous Poynting vector, i.e. $\langle \boldsymbol{S} \rangle = \frac{1}{2}Re(\boldsymbol{E} \times \boldsymbol{H}^*)$, in which $\boldsymbol{H}^*$ denotes the complex conjugate of the magnetic field associated with the propagating mode.

Figure 1f shows the spatial profile of the Purcell factor simulated near the band-edge of the glide-plane PCW at a point where the group index is $n_g$=100. For the calculation, we assume a circularly polarized dipole source, i.e. $\hat{n}_D = (\hat{x} \pm i\hat{y})/\sqrt{2}$. As shown in the Figure, the highest Purcell factors occur in the air holes, which is not compatible with epitaxial QDs. Nevertheless large Purcell enhancements up to 30 are available for QDs embedded in the line defect with optimal spectral tuning and spatial positioning close to the air holes, even when allowing for the dead zone at a distance ~40 nm from the holes [37].

A large number of devices were fabricated with a range of design parameters for the glide-plane PCWs giving photonic band edges between 880 nm and 940 nm. We identified the spectral position of the slow-light regions for a particular device by exciting the QD ensemble near one outcoupler and observing the spectrum transmitted through the device and collected from the opposite outcoupler. A clear cut-off was observed due to the adaptor's band gap (see figure S3 in Supplementary) which allowed the slow-light region to be identified by comparison with simulated band structures. Examples of the transmission measurements and the corresponding band structures that are determined for the central parts and adaptors are discussed in Figures S6 and S7 in Supplementary. The quantum dots were randomly positioned and had an inhomogeneous spread in emission frequencies. Multiple QDs were observed by exciting along the central PCW region, and those that were suitable for detailed study were selected according to their emission frequency relative to the photonic band edge of the particular device. The relative brightness observed at the outcoupler was used as an indicator of good coupling to the waveguide.

**Results: Non-chirally-coupled dot**

We first consider a non-chiral dot with the goal of establishing how large a Purcell factor can be obtained for these glide-plane structures. The Purcell factor depends on the precise position of the QD and its spectral position relative to the band edge. Figure 1f shows the expected variation of the Purcell factor for a QD emitting at the optimal frequency according to its location. In the experiments we searched for a bright QD emitting close a band edge and then carried out detailed lifetime measurements as the dot frequency was tuned.

The lifetime of the QD was measured using a mode-locked Ti:Sapphire laser and time-resolved photon counting methods. We performed the measurements in two different schemes of above-band non-resonant (808 nm) excitation and quasi-resonant (phonon-sideband) excitation as shown in Figures 2a and 2b, respectively. The above-band excitation method was sufficient for measurements with lifetimes $\gtrsim$ 200 ps, but for the fastest dots it was necessary to use the quasi-resonant method to eliminate the intra-dot relaxation processes through excited states that can slow down the decay. In these quasi-resonant excitation experiments, a laser detuning of around 2 meV was used to excite the phonon-side band of a QD line located in the slow-light region of the PCW structure (Figure 2b). This technique rapidly populates the target excited state $|X\rangle$ within the few-ps laser pulse[38,39], allowing us to reveal the true radiative lifetime[9].

Figures 2c and 2d presents data for a dot in a glide-plane PCW with a photonic band edge at around 911 nm. The dot could be tuned to the band edge by varying the bias, as shown by the photo-luminescence (PL) bias scan for the coupled QD shown in the inset to Figure 2d. Figure 2c compares the lifetime of the QD in the PCW (λ=911.2 nm at 0 V bias and -4.5 T magnetic field) to the QD ensemble lifetime and to the instrument response function. An exponential decay is observed, with a decay constant equal to the QD lifetime, namely 60±6 ps. The data were obtained under quasi-resonant excitation with the dot tuned to the slow-light region. In comparison with the QD ensemble in bulk, we extracted a 20±2 fold decay rate enhancement, i.e., a lifetime reduction from 1200 ps to 60±6 ps. The variation of decay rate enhancement for the same QD at different wavelengths, and for different schemes of quasi-resonant and above-band non-resonant excitations are shown in Figure 2e, indicating the impact of slow-light in the lifetime changes. As expected, the lifetime shortens as the QD is tuned to the band edge, and the lifetimes observed under quasi-resonant excitation are shorter than those for non-resonant excitation. The lifetime observed here is the shortest observed to date for a QD in a PCW.

The single-photon nature of the QD emission was confirmed by performing Hanbury Brown-Twiss (HBT) measurements. Figure S9-a in Supplementary shows the second order auto-correlation measurement for emission from the QD near the band edge when the lifetime is extremely fast. An anti-bunching dip was observed. The deconvolved fit (red curve) based on the measured 114 ps detector response time confirms the single photon nature of emission due to the strong antibunching dip ($g^2(0) = 0.11$). Furthermore, the HBT measurements for a dot slightly detuned from the band edge (at λ=912 nm) when the lifetime is ~200ps are shown in Figure S9-c. The result confirms the single-photon nature of emission (raw $g^2(0) = 0.16$) from the QD.

**Results: Chirally-coupled dot**

We now focus on a different dot that shows both strong chiral coupling and also a significant Purcell enhancement. Figure 3a shows simulations illustrating the dependence of the propagation direction of photons on the circular polarization of a dipole emitter positioned at a C-point. In the experiments, we identified such transitions by the oppositely polarized Zeeman components of the QD, as shown in the inset. The chiral coupling occurs when the QD is located in a chiral region of the waveguide. Figure 3b shows a simulated profile indicating chiral regions with opposite helicity by red and blue colours. We used Stokes parameters to calculate the degree of circular polarization of the internal electric field as given by $D = -2Im\{E_x E_y^*\}/I$, in which $I = |E_x|^2 + |E_y|^2$ is the intensity, $E_x$, and $E_y$ are in-plane components of the electric field, and * indicates complex conjugate. Two local waveguide points of *P1* and *P2* represent two extreme cases with zero directionality (linearly polarized light field) and pure circularly polarized light field (chiral point), respectively.

A comparison of Figures 3b and 1f shows that there are regions of high chirality near the air holes that overlap with regions with strong Purcell enhancement. It is a key feature of the glide-plane designs that they permit simultaneous achievement of high chirality and high Purcell factor. The simulations in Figure 3b indicate that it should be relatively easy to find dots with high chiral coupling in the glide-plane PCWs. This is illustrated in Figure 3c, where experimental results for 6 different QDs are shown. All of these dots show highly chiral coupling, with oppositely polarized Zeeman components coupling in opposite propagation directions, respectively.

While Figure 3b shows extensive areas of high chirality, not all of the chiral dots are expected to have large Purcell factors, as they need to be positioned in regions near the holes with high

Purcell factors. In addition, the emission frequency needs to coincide with the slow-light region. Figure 4 shows chiral coupling for one QD located near the slow-light region of the glide-plane PCW waveguide device with a band edge at 879 nm. The transmission measurements of the device (Figure S6-b in Supplementary), and the corresponding band structure simulations (Figure S6-c in Supplementary) consistently indicate the spectral location of the slow-light region at around 879 nm. Figure 4a illustrates the Zeeman components under ±3 T magnetic field of a QD that could be tuned from 874.4 nm at 0 V to 876.1 nm at 3 V as shown in Figure 4b. The strong directional emission for the Zeeman components shown in Figure 4a confirms the chiral coupling of the QD. Figure 4c compares the time-resolved PL at the two wavelengths. A clear decrease in lifetime from 420 ps at 874.4 nm to 255 ps at 876.1 nm is observed as the detuning from the band edge decreases from 4.6 nm to 2.9 nm. Compared with the ensemble lifetime, we calculate a 5±1 fold enhancement in the decay rate arising from the slow-light effects. This Purcell factor is not as large as for the non-chiral dot due to the inability to tune closer to the band edge. However, it is still substantially larger than for all previous measurements on chiral QDs, and shows the potential for obtaining very fast emission with samples that can be tuned closer to the band edge.

**Discussion**

The ability of an emitter-waveguide system to realize efficient directional coupling of spin-carrying photons can be characterized with the figure-of-merit (FOM) determined by

$$FOM = \frac{A_{chiral}}{A_{wg}} \cdot \frac{\Gamma_{tot}}{\Gamma_0} \cdot \frac{\Gamma_{wg}}{\Gamma_{tot}}, \qquad (2)$$

in which $\frac{A_{chiral}}{A_{wg}}$ is the fraction of the waveguide with >90% degree of circular polarization, $\frac{\Gamma_{tot}}{\Gamma_0}$ is the spontaneous emission rate enhancement calculated in the chiral area, and $\frac{\Gamma_{wg}}{\Gamma_{tot}}$ is the coupling efficiency, i.e. $\beta$-factor. The PCW system that we have developed reaches a FOM value of 10.8 ($\frac{A_{chiral}}{A_{wg}}$=36%, $\frac{\Gamma_{tot}}{\Gamma_0}$=30, $\beta$=99%) in theory (see Supplementary, Figure S2), and 1.8±0.36 in experimental demonstration where 5±1 Purcell enhancement is achieved for a chirally-coupled dot in the slow light region, clearly substantially outperforming previous demonstrations of chiral quantum-optical waveguide systems. A careful comparison of our PCW system with other waveguide systems is presented in Table 1. We have simulated all these structures to produce the values in Table 1. We also provide additional experimental results for the proposed slow-light waveguide system with different lattice constant to cover different spectral regions (see Supplementary, Figure S7). The data show that the glide-plane system gives better performance than all the other waveguide systems considered in the figure, namely nanobeams, glide-plane nanobeams, W1 waveguides, and topological valley Hall waveguides. The comparison results are consistent with other studies[34] and confirm the suitability of the glide-plane waveguide platform for chiral quantum optics. In Reference [34], the potential of topological waveguides are highlighted by introducing a different FOM.

**Conclusion**

We have shown that slowing down the speed of light in a nano-photonic waveguide can effectively enhance the light-matter interaction and map the spin qubit information that is carried by photons into pre-determined directions of the waveguide medium. We reported a record decay rate of 17±2 ns$^{-1}$ in a waveguide geometry, corresponding to a 20±2 fold Purcell-enhancement for a QD coupled to the slow-light regime of a waveguide. The very fast 60±6 ps lifetime was achieved using phonon sideband excitation to eliminate intradot relaxation processes that slow the decay. In addition, we demonstrated unidirectional spin-photon coupling in chiral points of a slow-light waveguide with a 5±1 fold Purcell enhancement, clearly outperforming previous measurements of Purcell-enhanced directional emission. These results demonstrate the excellent potential of QDs embedded within glide-plane waveguides for implementations of chiral quantum optics and realisation of on-chip optical quantum technologies

## Methods

A femtosecond pulsed Ti:Sapphire laser (Spectra-Physics Tsunami, Newport) with 82 MHz repetition rate was used for lifetime measurements. Non-resonant measurements were performed at 808 nm wavelength with 5 THz bandwidth and 0.2 ps pulse duration. Quasi-Resonant measurements were perform using 4f spectral filtering of the laser to ~6 ps pulse duration and ~160 GHz bandwidth. Additionally, an ultra-narrow band pass filter with FWHM < 0.55 nm bandwidth (935.4-0.45 OD5 Ultra Narrow Bandpass Filter, Alluxa) was angle-tuned into resonance with the QD under study, filtering the quasi-resonant laser out of the collected QD signal. In order to excite the QD, we applied additional weak non-resonant continuous-wave laser excitation (150 nW) together with the quasi-resonant laser (1.5 μW).

The QD emission was measured by a superconducting nanowire single photon detector (SNSPD - Single Quantum Eos), and the laser pulse repetition rate was measured by a photodiode. The pulses from the SNSPD and the photodiode were measured by a time-correlated photon counting card (Becker and Hickl SPC-130-EM). Two different SNSPD detectors were used with different instrument response functions. The fastest detector with 20 ps response time was used only for the quasi-resonant measurements with phonon sideband pumping presented in Figure 2. HBT measurements presented in Supplementary Figure S9 were performed with off-resonant excitation and with the slower detectors (convolved response time 114 ps).

For the nanophotonic design and modelling, finite-difference time-domain (FDTD) simulations were performed using commercial software (Ansys Lumerical FDTD). We used the bandstructure and dipole cloud analysis groups of the software to simulate the dispersion curves of our nanophotonic waveguide structures.

## Acknowledgements

This work was funded by the Engineering and Physical Sciences Research Council (EPSRC) UK Programme Grants EP/N031776/1 and EP/V026496/1. AJB acknowledges support from the EPSRC UK Quantum Technology Fellowship EP/W027909/1.

## Contributions

H.S. designed and simulated the photonic structures. E.C. and P.K.P. grew the quantum dot wafer. R.D. fabricated the photonic nanostructures and processed the QD wafer into diodes. H.S. and C.O.R. carried out the measurements with assistance from M.N.M.. A.J.B. contributed to the quasi-resonant measurements. D.J.H. contributed to the HBT measurements. A.M.F. and

**Figures**

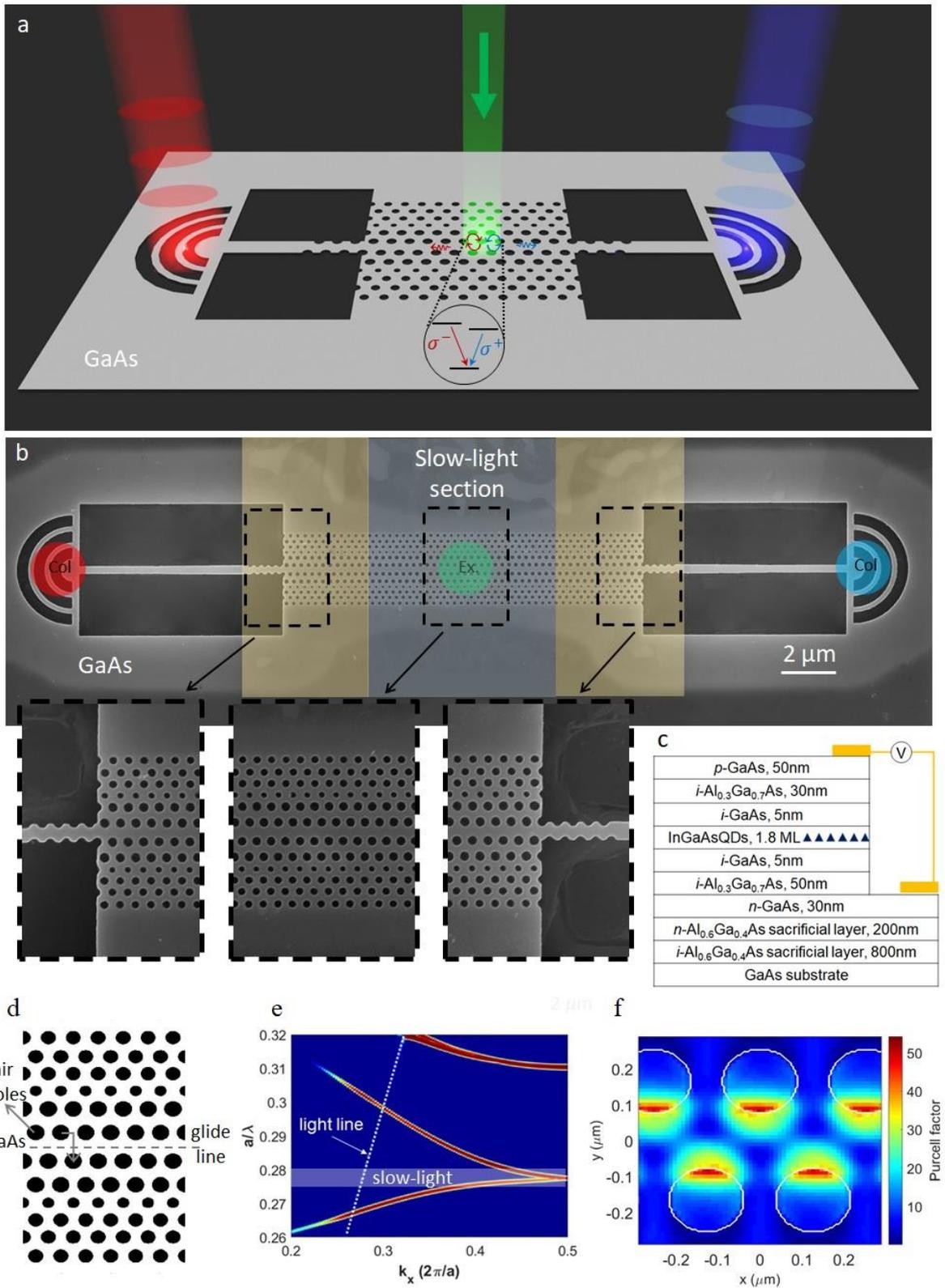

Figure 1: (a) Schematic of the device layout and working principle. (b) SEM image of the PCW device patterned on a GaAs wafer. (c) Schematic of the p-i-n GaAs diode structure with embedded InGaAs QDs. (d) Top-view of the slow-light section of the PCW with an up-down glide-plane symmetry. (e) Simulated band structure of the PCW. (f) Spatial profile of the Purcell factor inside the PCW at a frequency around the crossing point of the slow-light region.

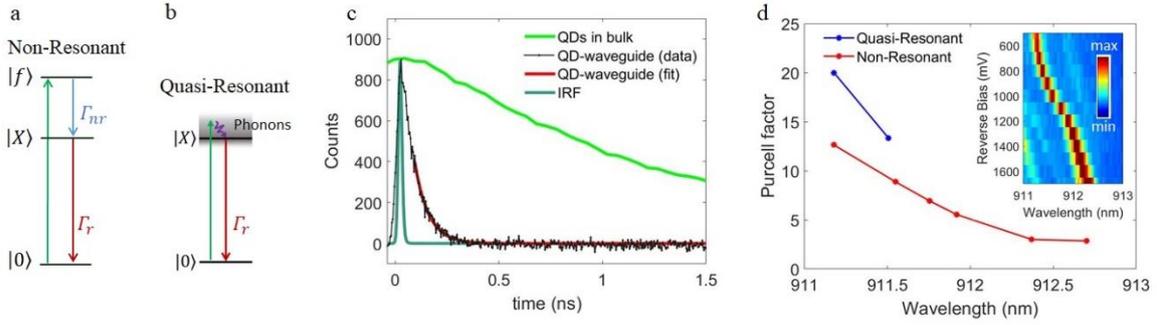

Figure 2: Time-resolved PL measurements for the non-chiral QD. (a) Non-resonant excitation scheme. (b) Quasi-resonant phonon-sideband excitation scheme. The grey box shows the spectral density of the phonon states by using a gradient fill so that the grey fades out as we get further away from the transition in energy. (c) Lifetime measurement for a tuned QD at the slow-light region (dark dots, V= 0 V, B= -4.5 T) under quasi-resonant phonon-sideband excitation. The solid red curve represents a single exponential fit. For the quasi-resonant excitation, a small laser detuning of 2 meV was used to excite the phonon-side band of the QD line. That gave a very fast (few *ps*) relaxation and thus revealed a radiative decay time of 60±6 *ps*. The light green line shows photoluminescence decay of the QD ensemble in bulk measured with non-resonant excitation. The dark green line represents the instrument response function (IRF, FWHM = 20 *ps*). (d) Wavelength dependency of the Purcell factor of the QD coupled to the slow-light PCW. The red and blue dots represent measured data under non-resonant above-band (808 nm) and phonon-sideband excitation schemes, respectively. The inset shows the PL scan for the coupled QD versus voltage.

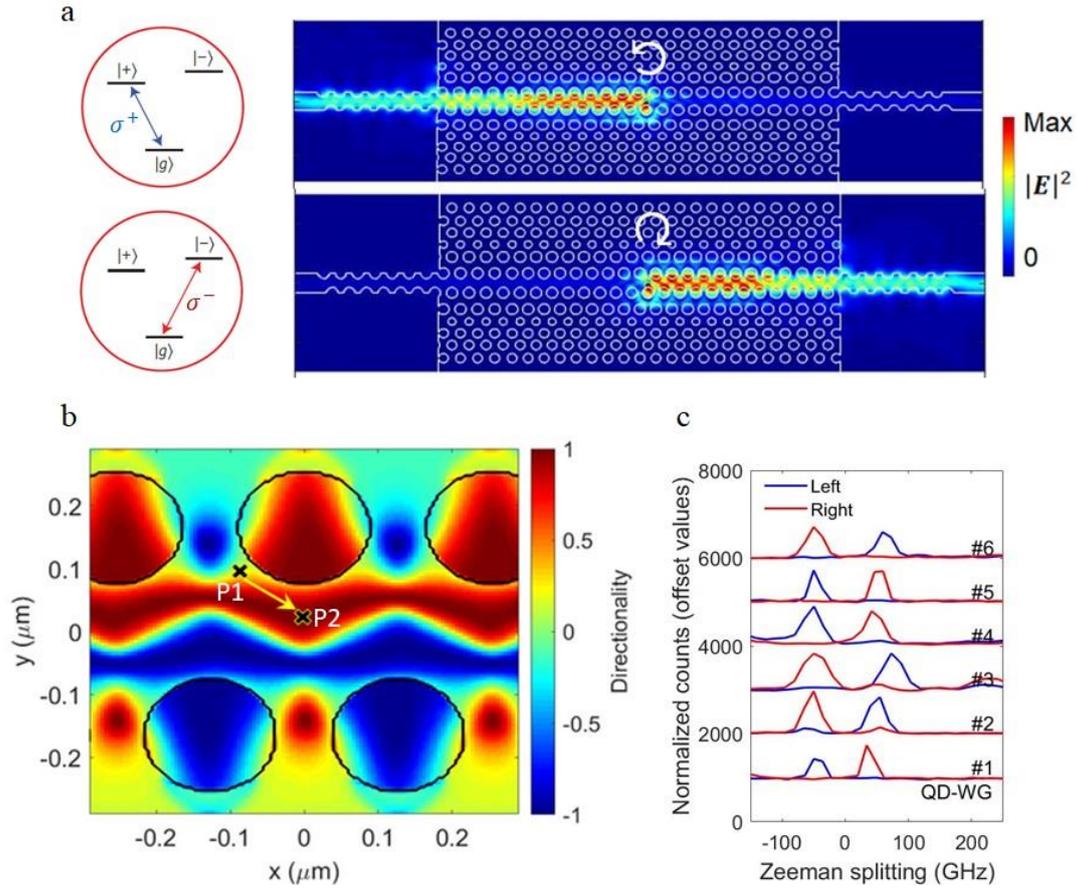

Figure 3: Chiral routing of spin-carrying photons. (a) Simulated electric field profile of the waveguide mode propagating along the line defect with a direction being determined by the spin state of the emitted photons from an embedded QD. The inset shows two transition dipole moments of $\sigma^+$ and $\sigma^-$ with orthogonal circular polarization. (b) Calculated directionality for circularly polarized dipoles as a function of position relative to the glide-plane waveguide structure. The degree of circular polarization of the internal electric field determined the chiral regions of the waveguide. Two local waveguide points of *P1* and *P2* represent extreme cases with zero directionality (linearly polarized light field) and pure circularly polarized light field (chiral point), respectively. (c) Optical measurement for QDs located in 6 different waveguides. The red and blue lines represent the photon collection from the left and right out-couplers, respectively. The Zeeman components of the QDs in an out-of-plane magnetic field of B= 3 T have opposite circular polarizations. The results indicate a strong chiral coupling of QDs into the waveguide mode.

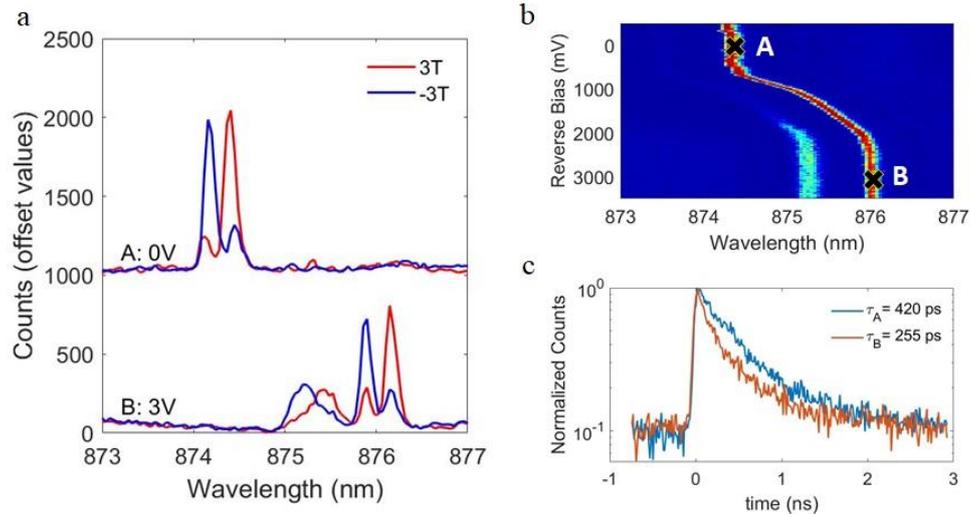

Figure 4: Purcell-enhanced chirally-coupled QD. (a) Chiral measurements of a QD near the slow-light region. The PL was collected from the right outcoupler of the device. Strong chirality was observed using magnetic field at 3T (red line) and -3T (blue line) for both 0 V bias (A point), and 3 V bias (point B). The broad line that emerges below the QD at 3V might be a charge state with short lifetime or another QD. (b) PL scan for the chirally coupled QD. (c) Lifetime measurement for the QD at point *A* (red) and point *B* (blue) using non-resonant excitation pulse laser at 808 nm. The fitted decay time decreased from 420 ps to 255 ps on red-shifting the QD emission line from 874.4 (point *A*) to 876.1 nm (point *B*).

Table 1: Comparison of chiral quantum optical waveguide systems including nanobeam (NB), W1-PC, glide-plane nanobeam (GPN), glide-plane waveguide (GPW), and topological Valley-Hall waveguide. The figure of merit (FOM) is defined in equation (2).

| Device | chiral area (C > 90%) | $\beta$-factor | Purcell-factor | FOM |
|---|---|---|---|---|
| NB | 28% | 83% | 1.1 | **0.31** |
| W1 | 0.8% | 99% | 2.8 | **0.02** |
| GPN | 28% | 83% | 2.2 | **0.61** |
| GPW | 36% | 99% | 30 | **10.8** |
| Topological Valley-Hall | 21% | 57% | 0.68 | **0.14** |